\documentclass[twocolumn,showpacs,preprintnumbers,amsmath,amssymb,superscriptaddress,prl]{revtex4}

\usepackage{graphicx} 
\usepackage{dcolumn}
\usepackage{bm}

\begin{document}

\title{Non-topological solitons in field theories with kinetic
self-coupling}

\author{Joaquin Diaz-Alonso}
\affiliation{LUTH, Observatoire de Paris, CNRS, Universit\'e Paris
Diderot. 5 Place Jules Janssen, 92190 Meudon, France}
\affiliation{Departamento de Fisica, Universidad de Oviedo. Avda.
Calvo Sotelo 18, E-33007 Oviedo, Asturias, Spain}
\author{Diego Rubiera-Garcia}
\affiliation{Departamento de Fisica, Universidad de Oviedo. Avda.
Calvo Sotelo 18, E-33007 Oviedo, Asturias, Spain}

\date{\today}

\newcommand{\be }{\begin{equation}}
\newcommand{\bea }{\begin{eqnarray}}
\newcommand{\bh }{\begin{displaymath}}
\newcommand{\en }{\end{equation}}
\newcommand{\ena }{\end{eqnarray}}
\newcommand{\eh }{\end{displaymath}}

\begin{abstract}

We investigate some fundamental features of a class of non-linear
relativistic lagrangian field theories with kinetic self-coupling.
We focus our attention upon theories admitting static, spherically
symmetric solutions in three space dimensions which are
finite-energy and stable. We determine general conditions for the
existence and stability of these non-topological soliton
solutions. In particular, we perform a linear stability analysis
that goes beyond the usual Derrick-like criteria. On the basis of
these considerations we obtain a complete characterization of the
soliton-supporting members of the aforementioned class of
non-linear field theories. We then classify the family of
soliton-supporting theories according to the central and
asymptotic behaviors of the soliton field, and provide
illustrative explicit examples of models belonging to each of the
corresponding sub-families. In the present work we restrict most
of our considerations to one and many-components scalar models. We
show that in these cases the finite-energy static spherically
symmetric solutions are stable against charge-preserving
perturbations, provided that the vacuum energy of the model
vanishes and the energy density is positive definite. We also
discuss briefly the extension of the present approach to models
involving other types of fields, but a detailed study of this more
general scenario will be addressed in a separate publication.

\end{abstract}


\maketitle

The non-linear Born-Infeld (BI) field theory was originally
introduced to remove the divergence of the electron's self-energy
in classical electrodynamics \cite{born34}. The BI procedure
defines a new lagrangian for the electromagnetic field, as a given
function of the two field invariants, which reduces to the Maxwell
lagrangian in the low-energy limit and exhibits central-field
static solutions which are finite-energy and stable. This
procedure has been extended and widely used in many different
contexts in Theoretical Physics. As examples of these extensions
we mention: a) The generalization of the BI procedure to
non-abelian gauge fields and to higher space dimensions, suggested
by some results in the context of M theory fivebrane
\cite{soliton}. b) The search for self-gravitating soliton
solutions of Einstein equations \cite{gravisoliton}. c) The
description of dark energy in the context of Cosmology, as a gauge
field governed by a BI-like action, or as scalar fields with
several kinds of derivative self-couplings \cite{cosmo}. d) The
elaboration of chiral models with soliton solutions for the
phenomenological description of the nucleon structure, as an
alternative to the usual Skyrme model with a stabilizing quartic
term in the lagrangian \cite{skyrme}. Almost all of these
generalizations introduce modified lagrangians which are obtained
from the initial ones using the same BI prescription. Besides
historical reasons, this choice is supported by some nice
properties of the original and extended BI actions, as the
existence of finite-energy (electrostatic and dyon-like) solutions
\cite{chern99}, the \textit{exceptional} properties of wave
propagation (in the electromagnetic original version
\cite{boillat70}) or the improvement of the stability behavior. It
is clear that the physical motivations for studying field theories
inspired on the BI approach are many-fold. Consequently, it is of
considerable interest to establish under what conditions these
field theories admit stable particle-like solutions. The aim of
the present work is to determine these conditions and to use them
to characterize the set of models supporting the aforementioned
type of solutions. As we shall see, the alluded conditions define
exhaustively this set of models and allow for their explicit
determination. This provides a wide panoply of supplementary tools
in order to address the aforementioned problems and others.

We have first approached these questions in the simpler case of scalar
fields ($\phi(x^{\mu})$), with lagrangian densities defined as
\textit{arbitrary functions} of the kinetic term ($X = \partial_{\mu}\phi.
\partial^{\mu}\phi$). This is the natural restriction to the scalar case of
the more general problem outlined for electromagnetic fields, where the
general lagrangians can be defined as arbitrary functions of the field
invariants. Moreover, the analysis of this problem and its extension to the
case of many-components scalar fields, are basic steps for the later
generalization to the case of abelian and non-abelian gauge fields. However
the results for gauge models will be presented in separate publications
\cite{dr071}, \cite{dr072}.

The search for this kind of scalar models exhibiting static soliton
solutions in three space dimensions, by circumventing the hypothesis of the
Derrick theorem \cite{derrick64}, has been already partially discussed in an
old paper \cite{diaz83}. Here we go beyond these results by establishing the
general conditions for the existence of such solutions and performing a
general analysis of their linear stability, beyond the Derrick
(\textit{necessary}) criterion. Most of this analysis can be easily
generalized to other dimensions.

Let us start with the lagrangian density
\be
L=f(\partial_{\mu}\phi.\partial^{\mu}\phi),
\label{eq:(1)}
\en
where the function $f(X)$ is assumed to be \textit{continuous and derivable}
in the domain of definition ($\Omega$). For the purposes of the present
analysis we shall call ``class-1" field theories the models (\ref{eq:(1)})
for which $f(X)$ is defined and regular everywhere ($\Omega \equiv \Re$) and
``class-2" field theories those with $\Omega \subset \Re, 0\in \Omega$ and
$\Omega$ open and connected. For obvious physical reasons other models are
excluded. In all cases the associated field equations take the form of a
local conservation law (in what follows we denote as $\dot{f}(X)$ and
$\ddot{f}(X)$ the first and second derivatives of $f(X)$, respectively)
\be
\partial_{\mu}\left[\dot{f}(X)\partial^{\mu} \phi\right] = 0.
\label{eq:(2)}
\en
The energy density obtained from the canonical energy-momentum tensor is
\be
\rho = 2\dot{f}(X) \left( \frac{\partial \phi}{\partial t}\right)^{2}  -
f(X).
\label{eq:(3)}
\en
We require this energy density to vanish in vacuum and to be positive
definite everywhere. This imposes the following supplementary restrictions
on the Lagrangian density:
\bea
&f(0)=0 \hspace{0.2cm};\hspace{0.2cm} \dot{f}(X) \geq 0 \hspace{0.2cm}
(\forall X) \hspace{0.2cm}; \hspace{0.2cm}f(X)\leq 0 \hspace{0.2cm} (\forall
X \leq 0)&
\nonumber \\
&\frac{d}{dX}(\frac{f^{2}(X)}{X}) \geq 0 \hspace{0.2cm} (\forall X > 0).&
\label{eq:(4)}
\ena
We shall call ``admissible" the field models satisfying these requirements.
For static spherically symmetric (\underline{SSS}) solutions $\phi(r)$,
Eq.(\ref{eq:(2)}) can be integrated once, leading to
\be
r^{2}\phi^{'}\dot{f}(-\phi^{'2}) = \Lambda,
\label{eq:(5)}
\en
where $\phi^{'} = d\phi/dr$, and $\Lambda$ is the integration constant. From
this equation and the conditions (\ref{eq:(4)}) we see that $\phi^{'}(r)$
(if unique) must be a monotonic function. Strictly speaking
Eq.(\ref{eq:(5)}) determines the field strength for $r>0$ only. If we
substitute the solutions of this equation in (\ref{eq:(2)}) we obtain a
Dirac $\delta$ distribution of weight $4\pi\Lambda$. We can then identify
this parameter as the ``source point charge" associated with the scalar SSS
solution (alternatively, if $\phi^{'}(r)$ is asymptotically coulombian,
$\Lambda$ can be interpreted as the total scalar charge, continuously
distributed in space with a density $\vec{\nabla}^{2}\phi$, in analogy with
the definitions in the non-linear BI models \cite{born34}).

The total energy of the SSS solutions obtained by integration of
(\ref{eq:(3)}) reads
\be
\varepsilon(\Lambda) = -4\pi\int_{0}^{\infty}r^{2}f(-\phi^{'2}(r,\Lambda))dr
= \Lambda^{3/2}\varepsilon(\Lambda=1),
\label{eq:(6)}
\en
where the last equality is a consequence of the invariance of the solutions
of Eq.(\ref{eq:(2)}) under the scale transformations $\phi(\vec{r},t)
\rightarrow \lambda^{-1}\phi(\lambda\vec{r},\lambda t)$. The convergence of
this integral depends on the behavior of the SSS field strength at the
origin and at $r\rightarrow \infty$. If we assume a power law expression
($\phi^{'}(r) \sim r^{q}$) in both cases \footnote[1]{This assumption
excludes transcendent asymptotic behavior of the field strength as, for
example, exponential damping at infinity, but our conclusions will not be
affected by this restriction. In fact such models belong to the case B-3
defined below.}, it follows from Eq.(\ref{eq:(5)}) that the integrand in
(\ref{eq:(6)}) behaves as $r^{2}f(-\phi^{'2}(r)) \sim \phi^{'}(r) \sim
r^{q}$ and thus the convergence of the energy integral requires $q > -1 $
when $r \sim 0$ and $q < -1$ as $r \rightarrow \infty$. Moreover, the
parameter $q$ determines the behavior of the lagrangian density $f(X)$
around the values of $X=-\phi^{'2}(r)$ in the limits of the integral. This
allows to obtain supplementary conditions to be imposed on this function in
order to have finite-energy SSS solutions. In this way the discussion of the
different possible behaviors of the solutions at $r \sim 0$ and as $r
\rightarrow \infty$ leads to a classification of all admissible models
supporting this kind of solutions (see reference \cite{dr072} for more
details). When $r \sim 0$ we can distinguish three subcases (see Fig.1). If
$0>q>-1$ (\underline{case A-1}) the lagrangian density shows a vertical
parabolic branch as $X \rightarrow -\infty$ and the field strength of the
SSS solution diverges at the origin, but the integral of energy converges
there and the field potential $\phi(r)$ remains finite. If $q=0$
(\underline{case A-2}) the field strength of the SSS solution remains finite
at the origin ($\phi^{'}(r \sim 0) \sim C - \beta r^{\sigma}$), and the
slope of the lagrangian density diverges at $X = -\phi^{'2}(0) = -C^{2}$.
The lagrangian $f(X)$ itself can take a negative finite value at this point
(for $\sigma > 2$; the scalar version of the BI model corresponds to $\sigma
= 4$) or show a vertical asymptote there (for $\sigma \leq 2$). The case
$q>0$ (\underline{case A-3}) must be discarded. Indeed, if $2 \geq q > 0$
the lagrangian density diverges in vacuum and if $q > 2$ the energy density
becomes negative around $X = 0$.

\begin{figure}
\includegraphics[width=8.6cm,height=4.3cm]{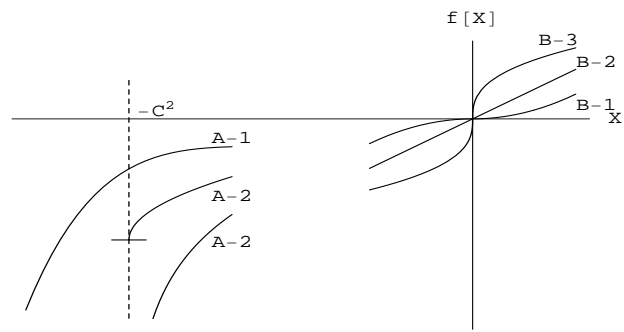}
\caption{\label{fig:epsart} Different possible behaviors of the
admissible lagrangians with finite-energy static central field
solutions.}
\end{figure}

Concerning the asymptotic behavior of the solutions we can also distinguish
three subcases (see Fig.1). When $-2 < q < -1$ (\underline{case B-1}) the
slope of the lagrangian density vanishes around $X \sim 0$. The definiteness
of the lagrangian around the origin restricts the admissible values of $q$
to the rational numbers, by means of the relation $q = \frac{4 +
2\Sigma}{\Sigma - 4}$, where $\Sigma = N_{1}/N_{2}$ is the irreducible ratio
of two odd naturals such that $N_{1} < N_{2}$. For $q = -2$ (\underline{case
B-2}) the lagrangian density behaves as the D'Alambert lagrangian around $X
\sim 0$ ($f(X \rightarrow 0^{\pm}) \sim X$) and the field strength is
asymptotically coulombian. When $q < -2$ (\underline{case B-3}) the slope of
the lagrangian diverges at $X \sim 0$, but the lagrangian itself remains
well defined there if the admissible values of $q$ are restricted to the set
of rational numbers through the formula $q = 2\frac{\Sigma + 1}{\Sigma -
1}$, where now $\Sigma = N_{1}/N_{2}$ is the irreducible ratio between an
even natural $N_{1}$ and an odd natural $N_{2} > N_{1}$.

Let us give three illustrative examples of admissible models belonging to
the different cases and showing soliton solutions (see Ref.\cite{dr072} for
a more detailed analysis of these models):
\bea
L_{1} &=&\frac{X}{2} + \lambda X^{a} \hspace{0.2cm};
\hspace{0.2cm} \lambda
>0, \hspace{0.2cm} a = \frac{odd}{odd} > \frac{3}{2}
\nonumber \\
L_{2} &=& \frac{(1+\mu^{2}X)^{\alpha} - 1}{2\alpha\mu^{2}}
\hspace{0.2cm};\hspace{0.2cm} 1/2 \leq \alpha < 1
\label{eq:(7)}\\
L_{3} &=&\frac{X^{\alpha}}{2(1+\mu^{2}X)^{a}}\hspace{0.01cm};
\hspace{0.01cm} 0<a = \frac{odd}{even}<1\hspace{0.01cm};
\hspace{0.01cm} \alpha = \frac{odd}{odd} > a+\frac{1}{2} \nonumber
\ena The first lagrangian ($L_{1}$) defines a two-parameter family
of class-1 field theories which fall into the cases A-1 and B-2
(field strength with an integrable singularity at the center and
coulombian asymptotic behavior). The constant $\lambda$ gives the
intensity of the self-coupling and the family reduces to the
D'Alambert model as $\lambda \rightarrow 0$. It can be generalized
to larger families of models with soliton solutions, including
lagrangians which are odd-powers series expansions in $X$ or odd
functions of $X$. The second lagrangian ($L_{2}$) defines a
two-parameter family of class-2 field theories, belonging to the
cases A-2 (finite limit of $f(X)$ at $X=-\phi^{'2}(0)$ and field
strength finite at the center) and B-2 (coulombian asymptotic
behavior). The value $\alpha = 1/2$ in this family corresponds to
the scalar version of the BI model. As the parameters $\mu
\rightarrow 0$ or $\alpha \rightarrow 1$ all the models converge
to the D'Alambert one, and the SSS solutions to the Coulomb field.
The third lagrangian ($L_{3}$) defines a three-parameter family of
class-2 field theories belonging to the cases A-2 (showing a
vertical asymptote at $X=-\phi^{'2}(0)$) and to the cases B-1, B-2
or B-3, depending on $ \alpha \gtreqqless 1$. In all cases the
strength of the field is finite at the origin ($\phi^{'}(0) =
1/\mu $) and behaves like $r^{-2/(2\alpha-1)}$ asymptotically.

Let us now address the question of the stability of the SSS
finite-energy solutions. The linear stability of these solutions
requires their energy to be a minimum against small perturbations
preserving the scalar charge of the soliton. At the first order
the (vanishing) modification of this charge by a small
perturbation $\delta\phi(\vec{r})$, assumed to be finite and
regular everywhere, takes the form \bea \Delta \Lambda = \int
d_{3}\vec{r} \vec{\nabla}.[\dot{f}(X_{0}) (\vec{\nabla}
\delta\phi)-
\nonumber \\
-2\ddot{f}(X_{0})(\vec{\nabla}\phi.\vec{\nabla}\delta\phi)
\vec{\nabla}\phi ] = 0, \label{eq:(8)} \ena where now $X_{0} =
-\phi^{'2}(r)$. The condition $\Delta \Lambda = 0$ imposes
restrictions on the behavior of the admissible perturbations at
$r=0$ and as $r\rightarrow\infty$. In particular, $\delta\phi$
must vanish asymptotically faster than the solution $\phi$
approaches its asymptotic value. In this way the perturbed fields
remain inside the space of functions defined by the prescribed
boundary conditions which determine uniquely the solutions.
Introducing the perturbed function in the integral of
Eq.(\ref{eq:(3)}) and expanding up to the second order, it is
easily seen that the first variation of the energy vanishes. This
is the necessary condition for the soliton energy to be an
extremum. The second variation reads \be \Delta_{2}\varepsilon =
\int d_{3}\vec{r} \left[ \dot{f}(X_{0}) (\vec{\nabla}
\delta\phi)^{2}
-2\ddot{f}(X_{0})(\vec{\nabla}\phi.\vec{\nabla}\delta\phi)^{2}\right],
\label{eq:(9)} \en which, owing to Eq.(\ref{eq:(8)}), converges
for any charge-preserving perturbation. The positivity of
$\Delta_{2}\varepsilon$, which is the sufficient condition for
linear stability, requires \be \dot{f}(X_{0}) +
2X_{0}\ddot{f}(X_{0}) = \frac{-2\Lambda}{r^{3}\phi^{''}(r)}
 > 0,
\label{eq:(10)}
\en
to be satisfied in all the range of values of $X=X_{0}$ covered by the
solution (the equality is obtained by deriving Eq.(\ref{eq:(5)}) with
respect to $r$). This condition is always fulfilled by admissible models
with finite-energy SSS solutions, owing to the monotonic character of
$\phi^{'}(r)$. We conclude that all
finite-energy SSS solutions of admissible models are linearly stable against
charge-preserving perturbations. We can perform a more detailed analysis of
the dynamics of the small perturbations ($\delta\phi(\vec{r},t)$) governed
by the linear equation
\bea
\vec{\nabla}.\left[\dot{f}(X_{0})\vec{\nabla}(\delta\phi) -2\ddot{f}(X_{0})
(\vec{\nabla}\phi.\vec{\nabla}\delta\phi)\vec{\nabla}\phi\right]-
\nonumber \\
-\frac{\partial}{\partial t}
\left(\dot{f}(X_{0}) \frac{\partial
\delta\phi}{\partial t}\right)=0, \label{eq:(11)}
\ena
obtained from the linearization of Eq.(\ref{eq:(2)}) around the SSS
solutions. Note that (\ref{eq:(11)}) also takes the form of a conservation
law. The general conditions imposed on the admissible models, besides the
knowledge of the behavior of the SSS solutions around $r = 0$ and as $r
\rightarrow \infty$, allow to perform the standard spectral analysis of
these linear problems, without explicit specification of the particular form
of the lagrangians \cite{dr072}. We summarize here the main conclusions of
this analysis, which can be deduced from the separation of space and time
variables in Eq.(\ref{eq:(11)}) together with the boundary condition
(\ref{eq:(8)}). There is, in all cases, a discrete spectrum of eigenvalues,
whose associated eigenfunctions are
mutually orthogonal and finite-norm, with respect to the scalar product
defined as the spatial integral of the products of the functions with
$\dot{f}(X_{0}(r))$ as kernel. The corresponding time dependence is
oscillatory. Any initial perturbation, bounded in the norm, remains bounded
as time evolves, confirming the linear stability.

All these results can be generalized to the case of $N$-components scalar
fields with a dynamics governed by lagrangian density functions of the form
\be
L(\phi_{i}, \partial_{\mu}\phi_{i}) = f\left(\sum_{i=1}^{N}
\partial_{\mu}\phi_{i}.\partial^{\mu}\phi_{i}\right),
\label{eq:(12)}
\en
where, as in the scalar case, $f(X)$ is a given \textit{continuous,
derivable and monotonically increasing} function. As already mentioned, the
analysis of this problem is also a necessary step in the generalization of
these methods to non-abelian gauge field theories. The field equations
associated with the lagrangians (\ref{eq:(12)}) take now the form of $N$
local conservation laws:
\be
\partial_{\mu}\left(\dot{f}(X)\partial^{\mu} \phi_{i}\right)  = 0,
\label{eq:(13)}
\en
where $X =\textstyle
\sum_{i=1}^{N}\partial_{\alpha}\phi_{i}.\partial^{\alpha}\phi_{i}$. For the
SSS solutions $\phi_{i}(r)$, these equations have $N$ first integrals of the
form
\be
r^{2}\phi_{i}^{'}\dot{f}\left(-\sum_{j=1}^{N}\phi_{j}^{'2}\right) =
\Lambda_{i},
\label{eq:(14)}
\en
where $\phi_{i}^{'} = d\phi_{i}/dr$ and $\Lambda_{i}$ are integration
constants. The canonical energy density reads
\be
\rho(x) = 2\dot{f}(X)\sum_{i=1}^{N}\left(\frac{\partial \phi_{i}}{\partial
t}\right)^{2} - f(X),
\label{eq:(15)}
\en
and is positive definite under the same conditions constraining the function
$f(X)$ in the one-component case (in fact the set of conditions
(\ref{eq:(4)}) defining ``admissibility" are assumed to hold also in this
case). In order to solve the system (\ref{eq:(14)}) we introduce the
functions $X_{i}(r) = -\phi_{i}^{'2}(r)$ and $X(r) =
\sum_{i=1}^{N}X_{i}(r)$. By squaring and adding Eqs.(\ref{eq:(14)}) we are
lead to
\be
r^{2}\sqrt{-X}\dot{f}(X) = \Lambda,
\label{eq:(16)}
\en
where $\Lambda = \sqrt{\sum_{i=1}^{N}\Lambda_{i}^{2}}$. We see that this
equation is formally identical to Eq.(\ref{eq:(5)}) and, if the function
$f(X)$ is the same in both cases, there is a one to one correspondence
between the solutions of the scalar case $\phi^{'}(r,\Lambda)$ and the
spheres $S_{\Lambda}$ of radius $\Lambda$ in the $N$-dimensional
$\Lambda_{i}$-space, associated with sequences of solutions of the
multiscalar case of the form
\be
\phi_{i}^{'}(r,\Lambda_{j}) =
\frac{\Lambda_{i}}{\Lambda}\phi^{'}(r,\Lambda).
\label{eq:(17)}
\en
The constants $\Lambda_{i}$ can now be identified as ``charges" associated
with the different components of the SSS field, in analogy with the BI and
the one-component scalar cases. The energy of these solutions, obtained from
the integration of (\ref{eq:(15)}), is the same as the energy of their
scalar counterparts, obtained from (\ref{eq:(6)}). There is a degeneracy in
$S_{\Lambda}$, obviously related to the rotational symmetry in the internal
space of the lagrangian (\ref{eq:(12)}). Moreover, the conditions
determining the multiscalar field models with finite-energy SSS solutions
coincide with those already discussed in the scalar case. Concerning the
conditions for stability of the solutions (\ref{eq:(17)}), the analysis of
the one-component case can be straightforwardly generalized to the present
situation \cite{dr072}. The final conclusion is that the multicomponent SSS
solitons are linearly stable against any perturbation preserving the scalar
charges $\Lambda_{i}$, if the associated one component solitons are linearly
stable (note also that transitions in the degeneration sphere $S_{\Lambda}$
are blocked by the charge conservation conditions).

Some additional comments on the general stability of these
\textit{non-topological} solitons are in order. Obviously, their
linear stability does not guarantee the conservation (or even a
proper definition) of the ``soliton identity" in presence of
strong ``external" fields. As is well known, in many examples of
\textit{topological} solitons their presence in any field
configuration can be detected through the existence of associated
discrete topological charges, which are conserved no matter the
intensity of external interactions. In some few cases (always in
one-space dimension) explicit exact many-soliton solutions have
been found, allowing the direct analysis of the dynamics of the
system in terms of interacting solitons \cite{scott73}. But a
satisfactory \emph{general} analysis of the interaction of
solitons with strong external fields in three space dimensions is
still lacking. However, there are some tentative approaches to
this question which have been developed in the framework of the
Born-Infeld model. We mention the method advanced in Ref.
\cite{chern98}, based on the use of the discontinuity of the field
strength at the center of \textit{static} BI solitons, as a marker
of the presence and location of the \textit{dynamic} soliton in
strong external fields. Since all our soliton solutions exhibit
similar central singularities, this procedure could be generalized
to the models considered here. But, in any case, the permanence of
these solitons in strong interactions remains a hypothesis which,
at best, is compatible with this method. On the other hand, in the
case of weak external fields linear stability \textit{does} imply
identity preservation. Consequently, linear stability is a basic
condition for the consistency of \textit{low-energy} calculations
of the interaction between solitons and weak fields (or between
distant solitons). The results of such calculations may be
interpreted in terms of particle-field (or particle-particle)
force laws and give also a first approach to the radiative
behavior in these processes \cite{chern99}.

As already mentioned, the present analysis has been generalized to
electromagnetic and non-abelian gauge fields of compact semisimple Lie
groups \cite{dr071}. The main result amounts to establish a correspondence
between any scalar model with stable SSS solutions of finite energy and
families of gauge field models exhibiting similar solutions (which can be
explicitly written in terms of the scalar ones). Conversely, the families so
defined exhaust the class of gauge field models supporting the
aforementioned kind of solitons.\\

\textbf{Acknowledgements}\\

We are grateful to Dr. A.R. Plastino for a careful reading of the
manuscript and very helpful comments.

\end{document}